\documentclass[prd,twocolumn,nofootinbib,superscriptaddress,amsmath,amssymb,groupedaddress]{revtex4-1}
\usepackage{mathtools}
\usepackage{graphics}
\usepackage{graphicx}
\usepackage{dcolumn}
\usepackage{bm}
\usepackage{dsfont} 
\usepackage{amsmath,amssymb}
\usepackage{verbatimbox}
\usepackage{hyperref}
\usepackage{tabularx}
\usepackage{epsf,epsfig}
\usepackage[normalem]{ulem}
\usepackage[usenames]{color}
\usepackage{multirow}
\usepackage{makecell}
\usepackage{diagbox}
\usepackage[abs]{overpic}
\allowdisplaybreaks
\hypersetup{
    colorlinks=true,
    linkcolor=blue,
    filecolor=magenta,      
    urlcolor=blue,
    citecolor=blue
}
\newcommand{\ISCO}{{\mbox{\tiny ISCO}}}
\newcommand{\EdGB}{{\mbox{\tiny EdGB}}}

\newcommand{\RD}{{\mbox{\tiny RD}}}

\newcommand{\ppE}{{\mbox{\tiny ppE}}}
\newcommand{\GR}{{\mbox{\tiny GR}}}
\newcommand{\GB}{{\mbox{\tiny GB}}}
\newcommand{\EH}{{\mbox{\tiny EH}}}

\definecolor{red(ncs)}{rgb}{0.77, 0.01, 0.2}

\usepackage{array}
\newcolumntype{C}[1]{>{\centering\let\newline\\\arraybackslash\hspace{0pt}}m{#1}}

\newcolumntype{C}[1]{>{\centering\arraybackslash}m{#1}}

\begin{document}

\title{Probing Einstein-dilaton Gauss-Bonnet Gravity with the inspiral and ringdown of gravitational waves}

\author{Zack Carson}
\author{Kent Yagi}

\affiliation{Department of Physics, University of Virginia, Charlottesville, Virginia 22904, USA}

\date{\today}


\begin{abstract}
Gravitational waves from extreme gravity events such as the coalescence of two black holes in a binary system fill our observable universe, bearing with them the underlying theory of gravity driving their process.
One compelling alternative theory of gravity -- known as Einstein-dilaton Gauss-Bonnet gravity motivated by string theory -- describes the presence of an additional dilaton scalar field coupled directly to higher orders of the curvature, effectively describing a ``fifth force" interaction and the emission of scalar dipole radiation between two scalarized black holes.
Most previous studies focused on considering only the leading correction to the inspiral portion of the binary black hole waveforms.
In our recent paper, we carried out inspiral-merger-ringdown consistency tests in this string-inspired gravity by including corrections to both the inspiral and ringdown portions, as well as those to the mass and spin of remnant black holes, valid to quadratic order in spin.
We here extend the analysis by directly computing bounds on the theoretical coupling constant using the full inspiral-merger-ringdown waveform rather than treating the inspiral and merger-ringdown portions separately. 
We also consider the corrections valid to quartic order in spin to justify the validity of black hole's slow-rotation approximation.
We find the quasinormal mode corrections to the waveform to be particularly important for high-mass events such as GW170729, in which the dilaton fields' small-coupling approximation fails without such effects included.
We also show that future space-based and multiband gravitational-wave observations have the potential to go beyond existing bounds on the theory.
The bounds presented here are comparable to those found in via the inspiral-merger-ringdown consistency tests.
\end{abstract}

\maketitle


\section{Introduction}\label{sec:intro}

On September 14, 2015, the Laser Interferometer Gravitational-wave Observatory (LIGO)  in Hanford and Livingston chirped with activity as they, for the first time ever, observed the iconic gravitational wave (GW) signal from the explosive coalescence of two black holes (BHs) 1.4 billion lightyears away.
Aptly named GW150914~\cite{GW150914} by the LIGO/Virgo Collaborations (LVC), this historic detection has ushered in an entirely new era of observational astrophysics, finally allowing us to probe the \emph{extreme gravity} regime of spacetime~\cite{Abbott_IMRcon2,Yunes_ModifiedPhysics,Berti_ModifiedReviewSmall}, where the fields are strong, non-linear, and highly dynamical.
GWs such as these carry multitudes of information across the universe regarding the local spacetime properties of the event, including clues highlighting the underlying theory of gravity driving the show~\cite{Gair:2012nm,Yunes:2013dva,Berti_ModifiedReviewLarge}.
For the past 100 years, Einstein's theory of general relativity (GR) has remained at its post as the prevailing theory of gravity, despite GW150914 and the following 10 events~\cite{GW_Catalogue} all being found to be consistent with his theory~\cite{TheLIGOScientific:2016pea,Abbott_IMRcon,Monitor:2017mdv,Abbott:2018lct}.
Even though the marvel of modern engineering that is the current LVC infrastructure~\cite{aLIGO} might not yet be sensitive enough to expose the subtle signs of a theory beyond GR, the next generation of ground- and space-based GW detectors~\cite{Ap_Voyager_CE,ET,LISA,B-DECIGO,DECIGO,TianQin} promise hefty sensitivity improvements across the GW frequency spectrum.
This may yet prove to finally be enough to study the traces of a new hidden theory of gravity describing our universe. 

For the past 100 years, GR has been put under the microscope, with countless observations and tests performed in a wide variety of spacetime environments, all ultimately finding agreement with Einstein's famous theory.
Observations on the solar-system scale where gravity is weak and approximately static~\cite{Will_SolarSystemTest}, or the strong-field, static observations of binary pulsar systems~\cite{Stairs_BinaryPulsarTest,Wex_BinaryPulsarTest}, even cosmological observations~\cite{Ferreira_CosmologyTest,Clifton_CosmologyTest,Joyce_CosmologyTest,Koyama_CosmologyTest,Salvatelli_CosmologyTest}, and extreme-gravity observations of GWs~\cite{Abbott_IMRcon2,Yunes_ModifiedPhysics,Berti_ModifiedReviewSmall,TheLIGOScientific:2016pea,Abbott_IMRcon,Monitor:2017mdv,Abbott:2018lct}, have all ultimately found results remarkably consistent with the predictions of GR.

Even with the substantial list of past observational success, we must continue to test GR.
While this theory still explains all of our gravitational observations, there still remains several open questions which could potentially be explained by alternative theories of gravity.
To give a few examples, the accelerated expansion of the universe due to dark energy~\cite{Jain:2010ka,Salvatelli:2016mgy,Koyama:2015vza,Joyce:2014kja}, the inconsistent galactic rotation curves due to dark matter~\cite{Famaey:2011kh,Milgrom:DarkMatter,Milgrom:2008rv,Clifton:2011jh,Joyce:2014kja}, the matter/anti-matter asymmetry in the current universe~\cite{Clifton:2011jh,Famaey:2011kh}, the inflationary period of the early universe~\cite{Joyce:2014kja,Clifton:2011jh,Famaey:2011kh,Koyama:2015vza}, or even the question of unifying GR and quantum mechanics~\cite{Clifton:2011jh,Joyce:2014kja,Famaey:2011kh,Milgrom:2008rv,Jain:2010ka,Koyama:2015vza} all remain open to this day.
Several modified theories of gravity have been proposed to date, many of which have been found to explain some of the open questions remaining.
Similar to the historical Newtonian description of gravity, these advanced theories could potentially reduce to GR in weak-gravity environments, and activate in the un-probed extreme-gravity spacetimes, such as outside binary BH mergers.

In particular, we consider an interesting classification of gravitational theories known as (massless) scalar-tensor theories, in which a massless scalar field is introduced.
Specifically, we focus our attention on a particular string-inspired scalar-tensor theory known as Einstein-dilaton Gauss-Bonnet (EdGB) gravity, where a dilaton scalar field is coupled to a quadratic curvature term in the action~\cite{Kanti_EdGB,Maeda:2009uy,Sotiriou:2013qea,Yagi:2015oca}, with coupling parameter $\alpha$.
With this new interaction in hand, BHs can become \emph{scalarized}~\cite{Campbell:1991kz,Yunes:2011we,Takahiro,Sotiriou:2014pfa,Berti_ModifiedReviewSmall,Prabhu:2018aun,Bakopoulos:2018nui,Antoniou:2017hxj,Antoniou:2017acq} (similar to conducting spheres becoming electrically charged), and a new fifth force interaction can be experienced between two such objects in a binary orbit.
Similar to analogous interactions found in nature (i.e. electromagnetic dipole radiation), such binary systems would decay faster than proposed by GR through additional scalar dipole radiation.

The current observational constraint found on the EdGB coupling parameter to date has been set to $\sqrt{\alpha}\lesssim2$ km~\cite{Kanti_EdGB,Pani_EdGB,Yagi_EdGB,Blazquez-Salcedo:2017txk,Witek:2018dmd,Nair_dCSMap,Yamada:2019zrb,Tahura:2019dgr}. Previous work on constraining EdGB gravity with GWs from BBH mergers mainly focused on looking at the correction in the inspiral due to the scalar dipole emission~\cite{Yunes_ModifiedPhysics,Nair_dCSMap,Yamada:2019zrb,Tahura:2019dgr}. BH quasinormal modes (QNMs) can also be used to probe this theory~\cite{Blazquez-Salcedo:2017txk}, while Ref.~\cite{Witek:2018dmd} estimated a rough bound on the theory from the dephasing due to the scalar field radiation computed via numerical relativity simulations. 
Additionally, see Ref.~\cite{Barausse:2016eii} where the authors found constraints on dipole emission with space-based detector LISA, as well as multiband observations.
See also a recent analysis by Ref.~\cite{Okounkova:2020rqw}, where the first numerical relativity model of an EdGB merger-ringdown waveform was presented, finding a coupling parameter constraint of $\sqrt{\alpha}\leq11$ km.

In this investigation, we probe EdGB gravity with GWs from BBH mergers by including both inspiral and ringdown corrections.
The former correction is computed using the commonly used \emph{parameterized post-Einstenian} (ppE) formalism~\cite{Yunes:2009ke}, in which generic amplitude and phase modifications are introduced into the inspiral GR waveform, and the mapping to EdGB is known~\cite{Yagi:2011xp,Yunes_ModifiedPhysics,Tahura:2019dgr}.
The latter corrections are computed with the EdGB corrections to the individual QNM ringing frequency and damping time found in Ref.~\cite{Blazquez-Salcedo:2016enn} (see also~\cite{Blazquez-Salcedo:2017txk,Blazquez-Salcedo:2016yka,Blazquez-Salcedo:2018pxo}). Moreover, we take into account EdGB corrections to the final mass and spin of the remnant BH as a function of the initial masses and spins, which can be estimated from corrections to the orbital energy and angular momentum found in Ref.~\cite{Ayzenberg:2014aka}.

This analysis follows closely along with that of Ref.~\cite{Carson_QNM_PRL} by the same authors, in which we considered the same EdGB corrections in GW signals (while template waveforms were still those in GR) and studied the inspiral-merger-ringdown (IMR) consistency tests of GR.
In particular, the EdGB coupling parameter $\alpha$ was allowed to vary until the inspiral and merger-ringdown portions of the waveform began to disagree with each other to a statistically significant degree.
At this point, the assumption of GR in the template waveform fails, and evidence of an emergent theory of gravity can be presented. 
The IMR consistency test was applied to a specific non-GR theory for the first time in Ref~\cite{Carson_QNM_PRL}.

In this paper, we directly estimate the measurement accuracy of $\alpha$ using the full IMR waveform, rather than treating the inspiral and merger-ringdown portions separately to look for their consistency. 
Moreover, BHs were assumed to be slowly-rotating, and corrections were derived up to quadratic order in spin in the previous analysis. 
Here, we also estimate how much higher-order corrections in spin may affect bounds on $\alpha$ by deriving corrections to quartic order in spin.

We now briefly summarize our findings.
With the EdGB corrections to the inspiral signal as well as the remnant BH mass, spin, and QNMs in hand, we derive current and projected future bounds on the EdGB coupling parameter $\alpha$.
As a first step calculation, we adopt the Fisher analysis~\cite{Cutler:Fisher} technique, which is known to agree well with Bayesian analyses for loud enough signals~\cite{Vallisneri:FisherSNR,Vallisneri:FisherSNR2}, such as GW150914~\cite{Yunes_ModifiedPhysics}.
We first consider four GW events, in order of increasing mass: GW170608~\cite{GW170608}, GW151226~\cite{GW151226}, GW150914~\cite{GW150914}, and GW170729~\cite{GW170729}.
We find that GW events detected during the O1/O2 runs by LVC detectors have varying success on the constraint of $\sqrt{\alpha}$ while varying the type of EdGB corrections introduced to the template waveform (inspiral only, axial or polar QNMs only, or both).
We find that for more massive events, the inclusion of corrections to the merger-ringdown are necessary in order to satisfy the small coupling approximation $16\pi\alpha^2/M^4 \ll 1$ for the total mass $M$ of a binary.
This stresses the need for the inclusion of merger-ringdown corrections to the template waveform, especially for more massive events where such contributions become important.
Further, we find that future GW150914-like events detected by CE~\cite{Ap_Voyager_CE}, LISA~\cite{LISA}, or the multiband combination of the two improve the constraints considerably, going beyond current bounds.
The results are similar to what were found in Refs.~\cite{Zack:Proceedings,Carson_multiBandPRL,Carson_multiBandPRD}, in which we only include corrections to the inspiral.
The resulting ``constraints" on $\sqrt{\alpha}$ found here are comparable to those from the IMR consistency test~\cite{Carson_QNM_PRL}.
We also find that higher-order terms in spin only affect the constraints $\sqrt{\alpha}$ by $1.6\%$ at most, justifying the slow-rotation approximation.

The outline of the paper is as follows.
We begin in Sec.~\ref{sec:EdGBoverall} discussing the theory of EdGB gravity. We also explain the resulting corrections we computed to the inspiral and merger-ringdown gravitational waveforms, as well as the remnant BH mass and spin.
Section~\ref{sec:testsOfGR} describes the Fisher analysis methodology used to derive bounds on EdGB gravity.
In Sec.~\ref{sec:results} we present and comment on our results, and investigate the effects of including higher-order BH spin corrections into the gravitational waveform.
Finally, we offer concluding remarks as well as a discussion of the various caveats to our work in Sec.~\ref{sec:conclusion}.
Throughout this paper, we have adopted geometric units such that $G=1=c$.


\section{Einstein-dilaton Gauss-Bonnet gravity}\label{sec:EdGBoverall}
In this section, we discuss the theoretical background relevant to the analysis.
We split this discussion into two distinct sections: beginning with a brief introduction into EdGB gravity, followed up with the various EdGB corrections to the gravitational waveform, including the inspiral, the quasinormal ringdown modes of the remnant BH, and finally the remnant BH's mass and spin.

\subsection{Theory}\label{sec:EdGB}

In this paper, we focus on the string-theory inspired EdGB theory of gravity.
In this particular theory, the ``dilaton" scalar field $\varphi$ is coupled to a quadratic curvature term in the action.
Correspondingly, the Einstein-Hilbert action is modified by the additional coupling term and the scalar field kinetic term~\cite{Kanti_EdGB,Maeda:2009uy,Sotiriou:2013qea}
\begin{equation}
\label{eq:EdGB}
S_\EdGB=\int d^4x \sqrt{-g} \left[f(\varphi) \mathcal{R}^2_{\GB} - \frac{1}{2} \nabla_\mu \varphi \nabla^\mu \varphi \right]\,,
\end{equation}
where $g$ is the determinant of the metric $g_{\mu\nu}$ and $\mathcal{R}_{\GB}$ is the curvature-dependent Gauss-Bonnet invariant given by
\begin{equation}
\mathcal{R}^2_\GB\equiv R_{abcd}R^{abcd}-4R_{ab}R^{ab}+R^2.
\end{equation}
$f(\varphi)$ is a function of $\varphi$ and some types of string theory effectively reduces to the correction in Eq.~\eqref{eq:EdGB} with an exponential coupling between the scalar field and $\mathcal{R}^2_\GB$. If one expands such a function about a fiducial value of the scalar field $\varphi_0$, the leading constant term does not contribute to the field equations since $\mathcal{R}^2_\GB$ is a topological invariant. Thus, the leading effect arises from a linear coupling, and in this paper, we consider\footnote{Additionally, refer to~\cite{Bakopoulos:2018nui,Antoniou:2017hxj,Antoniou:2017acq} for more general couplings.}
\begin{equation}\label{eq:linear-coupling}
f(\varphi) = \alpha \varphi\,,
\end{equation}
where $\alpha$ is the coupling parameter of the theory.

In scalar-tensor theories of gravity including EdGB gravity, compact objects can accumulate scalar monopole charges, which in turn source a scalar field.
This effect is naturally analogous to the classical effects of electric/mass/color charges sourcing the electric/gravitational/strong fields.
Pairs of such scalarized objects will then give rise to a new ``fifth force" interaction between them, altering their ensuing trajectories.
This effect is dependent on the internal structures of the compact objects, therefore violating the strong equivalence principle, one of the fundamental pillars of GR.
While two such compact objects orbiting each other in a binary system will decay under the emission of gravitational radiation (as predicted by GR)\footnote{Gravitational radiation is also modified from GR in EdGB gravity, though such an effect enters at higher order than the scalar dipole radiation in the binary evolution.}, the new scalar interaction will additionally induce scalar dipole radiation.
This effect will of course accelerate the coalescence process more than the predictions of GR estimate.

Scalar charges in EdGB gravity with a linear coupling as in Eq.~\eqref{eq:linear-coupling} only anchor to BHs~\cite{Campbell:1991kz,Yunes:2011we,Takahiro,Sotiriou:2014pfa,Berti_ModifiedReviewSmall,Prabhu:2018aun}, and not to other objects such as neutron stars~\cite{Yagi:2011xp,Yagi:2015oca}.
Such scalar charges $s$ depend on the BH's mass, spin, and the EdGB coupling parameter, and have been found to be~\cite{Berti_ModifiedReviewSmall,Prabhu:2018aun}
\begin{equation}
s_i=2\frac{\sqrt{1-\chi_i^2}-1+\chi_i^2}{\chi_i^2}\frac{\alpha}{m_i}.
\end{equation}
Here $\chi_i\equiv |\vec{S}_i|/m_i^2$ are the dimensionless spins of the $i$th BH with mass $m_i$ and spin angular momentum $\vec{S}_i$.

EdGB gravity may be treated as an effective field theory only if the correction $S_\EdGB$ to the action is much smaller than the Einstein-Hilbert action $S_\EH$.
Such an assumption allows one to neglect the higher-order curvature terms of order $\mathcal{O}(R^3)$, which correspond to cubic in the coupling parameter $\mathcal{O}(\alpha^3)$\footnote{$\varphi$ is of $\mathcal{O}(\alpha)$, and thus Eq.~\eqref{eq:EdGB} is of $\mathcal{O}(\alpha^2)$.}.
This approximation is known as the \emph{small coupling approximation}, and enforces the requirement that~\cite{Yunes_dcs,Yagi:2011xp}
\begin{equation}
\zeta \equiv \frac{16\pi\alpha^2}{M^4} \ll 1,
\end{equation}
for binaries with total mass $M\equiv m_1+m_2$.
If this inequality fails to be upheld, constraints on $\alpha$ are deemed to be invalid, as the assumption $S_\EdGB \ll S_\EH$ no longer holds.
Typically, constraints on the EdGB coupling parameter are presented for the quantity $\sqrt{\alpha}$, which has units of length (commonly in km).
Current constraints on this quantity have been found to be $10^7$ km from solar system observations~\cite{Amendola_EdGB}, and $\mathcal{O}(1\mathrm{km})$ from theoretical considerations, and observations of BH low-mass X-ray binaries, neutron stars, and GWs~\cite{Kanti_EdGB,Pani_EdGB,Yagi_EdGB,Witek:2018dmd,Nair_dCSMap,Yamada:2019zrb,Tahura:2019dgr}.

\subsection{Corrections to the gravitational waveform}\label{sec:corrections}
In this section, we describe the various corrections to the gravitational waveform taken into account in this analysis.
This includes corrections to the inspiral portion of the waveform, to the remnant BH's QNMs, and finally to the remnant BH's mass and spin predictions. We refer the readers to~\cite{Carson_QNM_PRL} for more details and actual expressions.
Finally, we point out a recent analysis in Ref.~\cite{Okounkova:2020rqw}, in which a numerical relativity binary BH merger-ringdown waveform in EdGB gravity was presented for the first time.
See also Ref.~\cite{Witek:2018dmd} for the scalar radiation during BH binary mergers in EdGB gravity, and Ref.~\cite{Okounkova:2019dfo} for a similar analaysis but in dynamical Chern-Simons gravity.

\subsubsection{Inspiral}\label{sec:inspiral}
In our analysis, we consider the commonly-used ppE formalism~\cite{Yunes:2009ke}\footnote{ppE phase corrections have a one-to-one correspondence to the inspiral corrections in the generalized IMRPhenom formalism~\cite{Abbott_IMRcon2} used by LVC~\cite{Yunes_ModifiedPhysics}.} to enact EdGB corrections to the inspiral gravitational waveform.
The ppE formalism allows one to modify the phase and amplitude of the GR waveform with generic parameterized corrections, in a theory-agnostic way.
The generalized ppE waveform can therefore be written as
\begin{equation}\label{eq:ppE}
\tilde{h}_{\text{ppE}}=A_{\GR}(f)(1+\alpha_{\text{ppE}} u^a)e^{i(\Psi_{\GR}(f)+\beta_{\text{ppE}} u^b)},
\end{equation}
where $\Psi_{\GR}$ and $A_{\GR}$ are the GR phase and amplitude respectively, $u=(\pi \mathcal{M} f)^{1/3}$ is the effective relative velocity of the compact objects with GW frequency $f$, and chirp mass $\mathcal{M}\equiv M \eta^{3/5}$, where $\eta\equiv m_1 m_2/M^2$ is the symmetric mass ratio.
Finally, the ppE parameters $\alpha_{\text{ppE}}$ $(\beta_{\text{ppE}})$ classify the magnitude of the amplitude (phase) modifications to the waveform which enter at $a$ $(b)$ power of velocity.
The parameters $a$ and $b$ are related to the \emph{post-Newtonian} (PN) order $n$ by $a=2n$ and $b=2n-5$, where terms entering the waveform at $n$PN order are proportional to $(u/c)^{2n}$ relative to the leading-order term.

In this investigation, we solely consider corrections to the GR waveform derived from the EdGB theory of gravity.
Such a theory affects the waveform amplitude at $-1$PN order ($a=-2$) with magnitude~\cite{Tahura_GdotMap}
\begin{equation}\label{eq:alpha_ppE}
\alpha_\ppE^{(\EdGB)}=-\frac{5}{192}\zeta\frac{(m_1^2 \tilde{s}_2-m_2^2\tilde{s}_1)^2}{M^4\eta^{18/5}},
\end{equation}
with $\tilde s_i = s_i m_i/\alpha$.
Similarly, the waveform phase is modified at $-1$PN order ($b=-7$) with magnitude~\cite{Yagi_EdGBmap}
\begin{equation}\label{eq:beta_ppE}
\beta_\ppE^{(\EdGB)}=-\frac{5}{7168}\zeta\frac{(m_1^2 \tilde{s}_2-m_2^2\tilde{s}_1)^2}{M^4\eta^{18/5}}.
\end{equation}
To have consistency with other EdGB corrections to be explained later, we only keep up to quadratic order in BH spins.
However, in Sec.~\ref{sec:4thOrdSpin} we consider EdGB corrections to the waveform up to $\mathcal{O}(\chi^4)$, comparing it to those found here.
For the remainder of this paper, corrections labeled ``inspiral" correspond to the addition of both phase and amplitude corrections to the GR inspiral waveform.
We note that Tahura \emph{et al.}~\cite{Tahura:2019dgr} showed that corrections to the GR amplitude is not as important as those in the phase, but we keep the former in this paper for completeness.

\subsubsection{Ringdown}\label{sec:mergerRingdown}
While the ppE formalism described above allows us to include EdGB corrections to the inspiral description of the waveform, we can additionally model corrections to the ringdown waveform.
As the orbits of the inspiraling BHs decay under the emission of gravitational radiation, they eventually become close enough to each other to enter plunging orbits, where a common horizon is formed as they merge together.
The remnant BH then relaxes down to its final state via the radiation of QNMs~\cite{Berti:2009kk}.
QNMs can be described by just two parameters: the ringdown frequency $f_\RD$ and the damping frequency $f_\text{damp}$~\cite{Berti:2005ys,Berti:2009kk}.
$f_\RD$ and $f_\text{damp}$ are described by the remnant BH's mass and spin $M_f$ and $\chi_f$ (from the BH no-hair theorem), which in turn only depend on $m_1$, $m_2$, $\chi_1$, and $\chi_2$ of the original BH binary system obtained through numerical relativity simulations~\cite{PhenomDII}.
See Ref.~\cite{Maselli:2019mjd} where similar corrections to the QNMs were made, and bounded by future observations of multiple GW events.
Additionally see Refs.~\cite{McManus:2019ulj,Cardoso:2019mqo} where they developed a new general formalism to map ringdown corrections to specific theories of gravity.

However, within the EdGB viewpoint of gravity, the QNMs additionally depend upon the coupling parameter $\zeta$.
In this analysis, we attempt to model corrections to the ringdown and damping frequencies $f_\RD$ and $f_\text{damp}$ up to first order in $\zeta$, like so
\begin{align}
f_{\RD} &=f_{\RD,\GR}+\zeta f_{\RD,\zeta} + \mathcal{O}(\zeta^2),\label{eq:fRD}\\
f_{\text{damp}} &=f_{\text{damp},\GR}+\zeta f_{\text{damp},\zeta} + \mathcal{O}(\zeta^2),\label{eq:fdamp}
\end{align}
where $f_{\RD,\GR}$ and $f_{\text{damp},\GR}$ are the GR QNM frequency predictions~\cite{PhenomDI,PhenomDII}, and $f_{\RD,\zeta}$ and $f_{\text{damp},\zeta}$ are the first order EdGB corrections. 
These quantities can be read off from~\cite{Blazquez-Salcedo:2016enn}\footnote{Reference~\cite{Blazquez-Salcedo:2016enn} follows a slightly different EdGB notation than considered here, beginning with the coupling parameter $\alpha$ in the action as well as their definition of $\zeta'$. The quantities can be mapped to our definitions by letting $\zeta' \rightarrow 4 \sqrt{\zeta}$.}, as can be found in~\cite{Carson_QNM_PRL}.

We consider both the $l=2$ axial QNMs, as well as the $l=2$ polar QNMs.
As discussed in Ref.~\cite{Blazquez-Salcedo:2016enn}, the QNM non-spinning components have been computed for both of these modes.
The spinning components of the axial modes were then obtained by adopting the null geodesic correspondence\footnote{In Refs.~\cite{Silva:2019scu,Glampedakis:2017dvb,Glampedakis:2019dqh}, the null geodesic correspondence was used to approximate corrections to rotating BHs as well.}~\cite{Yang:2012he} since such modes do not couple to the scalar field perturbation. The spinning components of the polar modes in EdGB gravity is currently unknown, though based on the claim in~\cite{Blazquez-Salcedo:2016enn}, we assume the polar modes have the same spin dependence as the axial modes to carry out a rough estimate on the latter.
In this analysis, we include EdGB corrections into the merger-ringdown waveform up to linear order in $\zeta$ and quadratic in $\chi_f$ using the above prescription, with corrections like so labeled as ``axial/polar QNMs".
See Sec.~\ref{sec:4thOrdSpin} for a demonstration of the inclusion of spin effects into the remnant BH QNMs, where we include corrections up to $\mathcal{O}(\chi^4)$, and also remove all spin effects.

\subsubsection{Remnant BH mass and spin}\label{sec:MfChif}
In addition to the direct waveform modifications displayed in the preceding sections, a post-merger remnant BH in EdGB gravity will settle down into a non-GR final mass and spin configuration, due to the increased levels of energy and angular momentum radiation.
This effect will also indirectly modify the gravitational waveform.
In GR, the final spin angular momentum of the post-merger BH can be roughly approximated to be the sum of the spin angular momentum of the initial BHs and the orbital angular momentum of a particle with mass $\mu=m_1 m_2/M$ orbiting about the remnant BH at the radius of the \emph{innermost stable circular orbit} (ISCO), $r_\ISCO$~\cite{Barausse:2009uz}.
More specifically, the full expression for a spin-aligned system is found to be~\cite{Buonanno:2007sv,Barausse:2009uz,Yunes_ModifiedPhysics}
\begin{equation}
\mu L_\text{orb}(M,\chi_f,r_\ISCO)=M (M\chi_f-a_s-\delta_m a_a),
\end{equation}
where $a_{s,a}\equiv(m_1\chi_1 \pm m_2\chi_2)/2$ are the symmetric/anti-symmetric combinations of spins, $\delta_m\equiv (m_1-m_2)/M$ is the weighted mass difference, and  $L_\text{orb}$ is the specific orbital angular momentum.
Similarly, the final mass of the remnant BH $M_f$ can be expressed in relation to the specific orbital energy $E_\text{orb}$ 
of a particle with mass $\mu$ orbiting at $r_\ISCO$ as
\begin{equation}
\label{eq:M_f}
M-M_f=\mu \left[1-E_\text{orb}(M_f,\chi_f,r_\ISCO) \right].
\end{equation}
Here $E_b\equiv 1-E_\text{orb}$ is equivalent to the binding energy of the particle. 

We here make an assumption that the above GR picture also holds in EdGB gravity and derive corrections to $M_f$ and $\chi_f$. To do so, we take into account the EdGB corrections to $E_\text{orb}$, $L_\text{orb}$ and $r_\ISCO$. Unfortunately, these expressions are not known to all orders in the BH spin. Thus, we use the expressions valid to quadratic order in spin presented in~\cite{Ayzenberg:2014aka}. In addition, there is a scalar interaction between two scalarized BHs, and thus Eq.~\eqref{eq:M_f} needs to be modified to
\begin{align}
M -M_f=&\mu \left[1-E_\text{orb}(M_f,\chi_f,r_\ISCO) \right. \nonumber \\
 & \left. - E_\text{scalar}(\mu,M,\chi_f,r_\ISCO,\zeta) \right].
\end{align}
Here 
\begin{equation}
E_\text{scalar}(\mu,M,\chi_f,r_\ISCO,\zeta)=\frac{\zeta}{\eta^2}\left( 1-\frac{\chi_f^2}{4} \right)\frac{M}{r}
\end{equation}
is the scalar interaction energy at $r_\ISCO$ between two particles with scalar charges~\cite{Stein:2013wza} taken up to quadratic order in spin.
See Sec.~\ref{sec:4thOrdSpin} for an investigation into higher-order spin effects up to $\mathcal{O}(\chi^4)$ for each EdGB correction considered here.

Similar to the merger-ringdown corrections to the QNM ringing and damping frequencies, we consider corrections to the remnant BH mass and spin to linear order in $\zeta$ and quadratic (and also quartic) in $\chi_f$.
The complete expressions for $M_f$ and $\chi_f$ can then be written as
\begin{align}
M_f &=M_{f,\GR}+\zeta M_{f,\zeta} + \mathcal{O}(\zeta^2),\label{eq:Mf} \\
\chi_f &=\chi_{f,\GR}+\zeta \chi_{f,\zeta} + \mathcal{O}(\zeta^2),\label{eq:chif}
\end{align}
where $M_{f,\GR}$ and $\chi_{f,\GR}$ are the GR predictions of the final mass and spin from the numerical relativity fits of Ref.~\cite{PhenomDII}, and $M_{f,\zeta}$ and $\chi_{f,\zeta}$ are the resulting EdGB corrections at first order in $\zeta$.


\section{Fisher Analysis}\label{sec:testsOfGR}
In this section, we provide a brief introduction to the analysis methods utilized in this paper.
We estimate the maximum size $\sqrt{\alpha}$ can take under a GW observation of a binary BH merger, while still remaining consistent within the detector noise. 

To this end, we employ the Fisher analysis techniques described in Refs.~\cite{Cutler:Fisher,Poisson:Fisher,Berti:Fisher,Yagi:2009zm}. 
A more comprehensive Bayesian analysis can be used instead to give more accurate results, yet is significantly more computationally expensive. For loud enough events however, the two have been shown to agree well~\cite{Vallisneri:FisherSNR,Vallisneri:FisherSNR2}. 
For example, the difference in the bounds on the leading PN correction to the inspiral portion of the waveform between Fisher and Bayesian analyses are small~\cite{Yunes_ModifiedPhysics} for GW150914. Similarly, regarding the inspiral-merger-ringdown consistency tests, Fisher results agree well with those from Bayesian analyses for GW150914~\cite{Carson_QNM_PRL}.

In the following investigation, we utilize the non-precessing, sky-averaged \emph{IMRPhenomD} waveform~\cite{PhenomDI,PhenomDII} together with the EdGB corrections described in Sec.~\ref{sec:EdGBoverall}.
Namely, the inspiral corrections to the amplitude and phase in Eqs.~\eqref{eq:ppE}--\eqref{eq:beta_ppE} are added to the inspiral portion of the IMRPhenomD waveform. The ringdown and damping frequencies in the merger-ringdown portion of the waveform is modified as Eqs.~\eqref{eq:fRD}--\eqref{eq:fdamp}. The mass and spin of the remnant BH entering in these frequencies needs to be related to the initial masses and spins, which are modified according to Eqs.~\eqref{eq:Mf}--\eqref{eq:chif}.
Thus, the gravitational waveform non-GR effects are all parameterized into the single EdGB parameter $\zeta$ (or $\alpha$).
The resulting template waveform consists of 
\begin{equation}
\label{eq:parameters}
\theta^a=\left( \ln{\mathcal{A}},\phi_c,t_c,\mathcal{M},\eta,\chi_s,\chi_a,\zeta \right),
\end{equation}
where $\mathcal{A}\equiv\frac{\mathcal{M}_z^{5/6}}{\sqrt{30}\pi^{2/3}D_L}$ is a frequency-independent part of the amplitude in Fourier space with redshifted chirp mass $\mathcal{M}_z\equiv\mathcal{M}(1+z)$ and redshift $z$, $D_L$ is the luminosity distance, $\chi_{a,s}\equiv(\chi_a \pm \chi_2)/2$ are the symmetric and anti-symmetric combinations of dimensionless spins, and $\phi_c$ and $t_c$ are the coalescence phase and time.

Let us now explain how one can obtain posterior probability distributions on template parameters $\theta^a$ using a Fisher analysis.
With the assumption that both the prior and template parameter distributions are Gaussian~\cite{Cutler:Fisher,Poisson:Fisher}\footnote{A Bayesian analysis can utilize more accurate probability distributions, such as uniform.}, the root-mean-square errors on parameters $\theta^i$ can be found to be
\begin{equation}
\Delta\theta^i=\sqrt{\tilde{\Gamma}_{ii}^{-1}},
\end{equation}
where $\tilde{\bm{\Gamma}}$ is the effective Fisher matrix given by
\begin{equation}
\tilde{\Gamma}_{ij}=\Gamma_{ij}+\frac{1}{(\sigma_{\theta^i}^{(0)})^2}\delta_{ij}, 
\end{equation}
while the \emph{Fisher information matrix} can be written as
\begin{equation}
\Gamma_{ij}\equiv \left( \frac{\partial h}{\partial \theta^i} \Bigg| \frac{\partial h}{\partial \theta^j} \right).
\end{equation}
$\sigma_{\theta^i}^{(0)}$ is the root-mean-square error on the Gaussian prior for the $i$th parameter.
In the above expression, the notation $(a|b)$ represents the inner product weighted by the detector noise spectral density $S_n(f)$
\begin{equation}
(a|b)\equiv2\int^{f_\text{high}}_{f_\text{low}}\frac{\tilde{a}^*\tilde{b}+\tilde{b}^*\tilde{a}}{S_n(f)}df,
\end{equation}
where $f_\text{high,low}$ represent the detector-dependent high and low cutoff frequencies, as are tabulated and described in Ref.~\cite{Carson_multiBandPRD}.
Finally, if one wishes to combine the observations from multiple detectors with Fisher matrices $\bm{\Gamma}^\text{A}$ and $\bm{\Gamma}^\text{B}$, the resulting effective Fisher matrix can be found to be
\begin{equation}
\tilde{\Gamma}^\text{tot}_{ij}=\Gamma^\text{A}_{ij}+\Gamma^\text{B}_{ij}+\frac{1}{(\sigma_{\theta^i}^{(0)})^2}\delta_{ij}.
\end{equation}
We adopt Gaussian priors corresponding to $|\phi_c|\leq\pi$, and $|\chi_{s,a}|\leq 1$.

\begin{figure}[htb]
\begin{center}
\includegraphics[width=\linewidth]{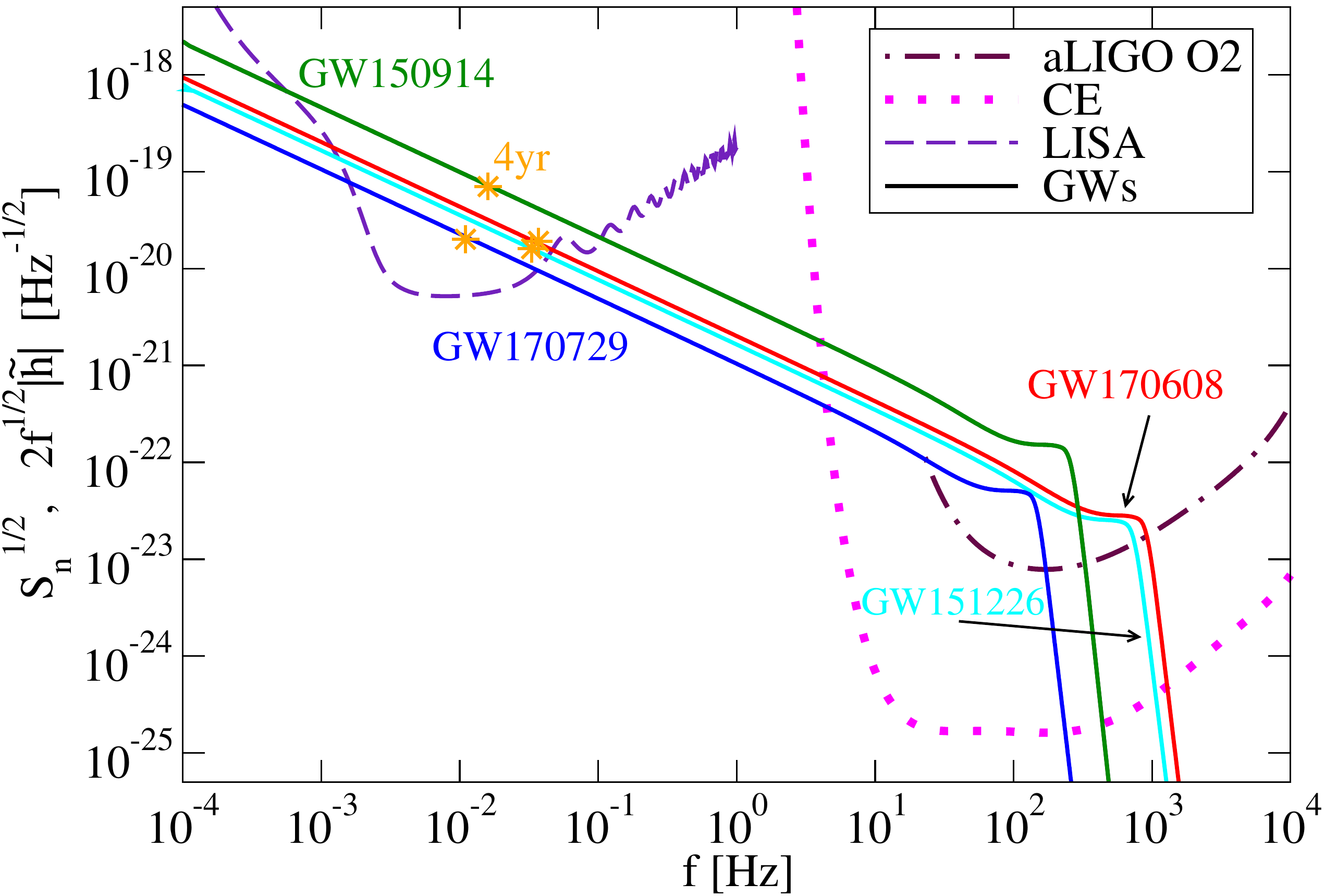}
\caption{
Sensitivities $\sqrt{S_n(f)}$ of the gravitational-wave interferometers aLIGO O2, CE, and LISA considered in this analysis.
We additionally display the characteristic amplitudes $2\sqrt{f}|\tilde h(f)|$ for GW events GW170608, GW151226, GW150914, and GW170729 with 4 years prior to merger displayed as orange stars.
}\label{fig:detectors}
\end{center}
\end{figure}

For the remainder of this analysis, we consider the following GW detectors, with sensitivities shown in Fig.~\ref{fig:detectors}.
We consider the current-generation LIGO/Virgo 2nd observing run (O2)~\cite{aLIGO} detector, as well as the future-planned third generation Cosmic Explorer (CE)~\cite{Ap_Voyager_CE} detector with a $\sim100$ times improvement in the frequency range $1-10^4$ Hz, and finally the future space-based detector LISA~\cite{LISA}, with advanced sensitivity in the mHz regime $10^{-4}-1$ Hz.
For the above detectors, $f_\text{low}$ is found to be $23$ Hz, $1$ Hz, and $f_\text{4yrs}$ respectively, where~\cite{Berti:2005ys}
\begin{equation}
f_\text{4yrs} = 1.09\times10^{-2}~\mathrm{Hz} \left( \frac{\mathcal{M}}{28 M_\odot} \right)^{-5/8}
\end{equation}
is the frequency $4$ years prior to merger.
Similarly, $f_\text{high}$ is found to be $1$ Hz for LISA, and 4,000 Hz for O2 and CE, such that the GW spectrum is sufficiently small compared to the detector sensitivity at $f_\text{high}$.

\renewcommand{\arraystretch}{1.2}
\begin{table*}
\centering
\addvbuffer[12pt 8pt]{\begin{tabular}{c | c c c c cc}
Event & $M$ ($M_\odot$) & $\mathcal{M}$ ($M_\odot$) & $\eta$ & $\chi_1$ & $\chi_2$ & $D_L$ (Mpc)\\
\hline
GW170608~\cite{GW170608} & 19.0 & 7.92 & 0.233 & 0.5 & $-0.66$ & 340\\
GW151226~\cite{GW151226} & 21.7 & 8.89 & 0.226 & 0.5 & $-0.36$ & 440\\
GW150914~\cite{GW150914} & 65.0 & 28.1 & 0.247 & 0.32 & $-0.44$ & 410\\
GW170729~\cite{GW170729} & 83.5 & 36.2 & 0.249 & 0.60 & $-0.57$ & 2,900\\
\end{tabular}}
\caption{List of binary BH GW events investigated in this analysis, including the most and least massive events yet detected (GW170729 and GW170608 respectively), along with their total mass $M$, chirp mass $\mathcal{M}$, symmetric mass ratio $\eta$, BH spins $\chi_{1,2}$, and the luminosity distance $D_L$ used in this paper.
}\label{tab:events}
\end{table*}

We consider four binary BH GW coalescence events in our analysis, ranging from small to large total mass $M$.
In order of increasing total mass, we have chosen GW170608~\cite{GW170608}, GW151226~\cite{GW151226}, GW150914~\cite{GW150914}, and GW170729~\cite{GW170729} with total masses $M$, chirp masses $\mathcal{M}$, symmetric mass ratios $\eta$ and luminosity distances ($D_L$) tabulated in Table~\ref{tab:events}.
Specifically, we perform Fisher analyses to estimate the extraction uncertainty on the EdGB parameter $\zeta$ using each detector and GW event considered above.
We choose the fiducial values of $\chi_s=\chi_a=\phi_c=t_c=0$.
We also choose $\zeta=0$ (GR), effectively making the resulting root-mean-square error on $\zeta$ to indicate the amount of non-GR ``fuzziness" one can expect the parameter to observe while still remaining consistent with GR, within the detector noise.

Additionally, we investigate the effect of each type of EdGB correction present in the template waveform: inspiral and ringdown effects.
In particular, we consider the following five cases in which we perform a Fisher analysis:
\begin{enumerate}
\item \textit{Inspiral:} EdGB corrections only in the inspiral waveform.
\item \textit{Axial QNMs:} EdGB corrections only in the ringdown waveform for the case of purely axial QNMs.
\item \textit{Polar QNMs:} Same as 2 but with polar QNMs.
\item \textit{Inspiral+Axial QNMs:} Combination of 1 and 2, with corrections to both the inspiral and ringdown portions.
\item \textit{Inspiral+Polar QNMs:} Same as 4 but for polar QNMs.
\end{enumerate}
We include remnant BH mass and spin corrections within only the latter four cases listed above.
Within each of the above listed cases, we compare the results from each detector and event considered.

\begin{figure}[htb]
\begin{center}
\includegraphics[width=.9\linewidth]{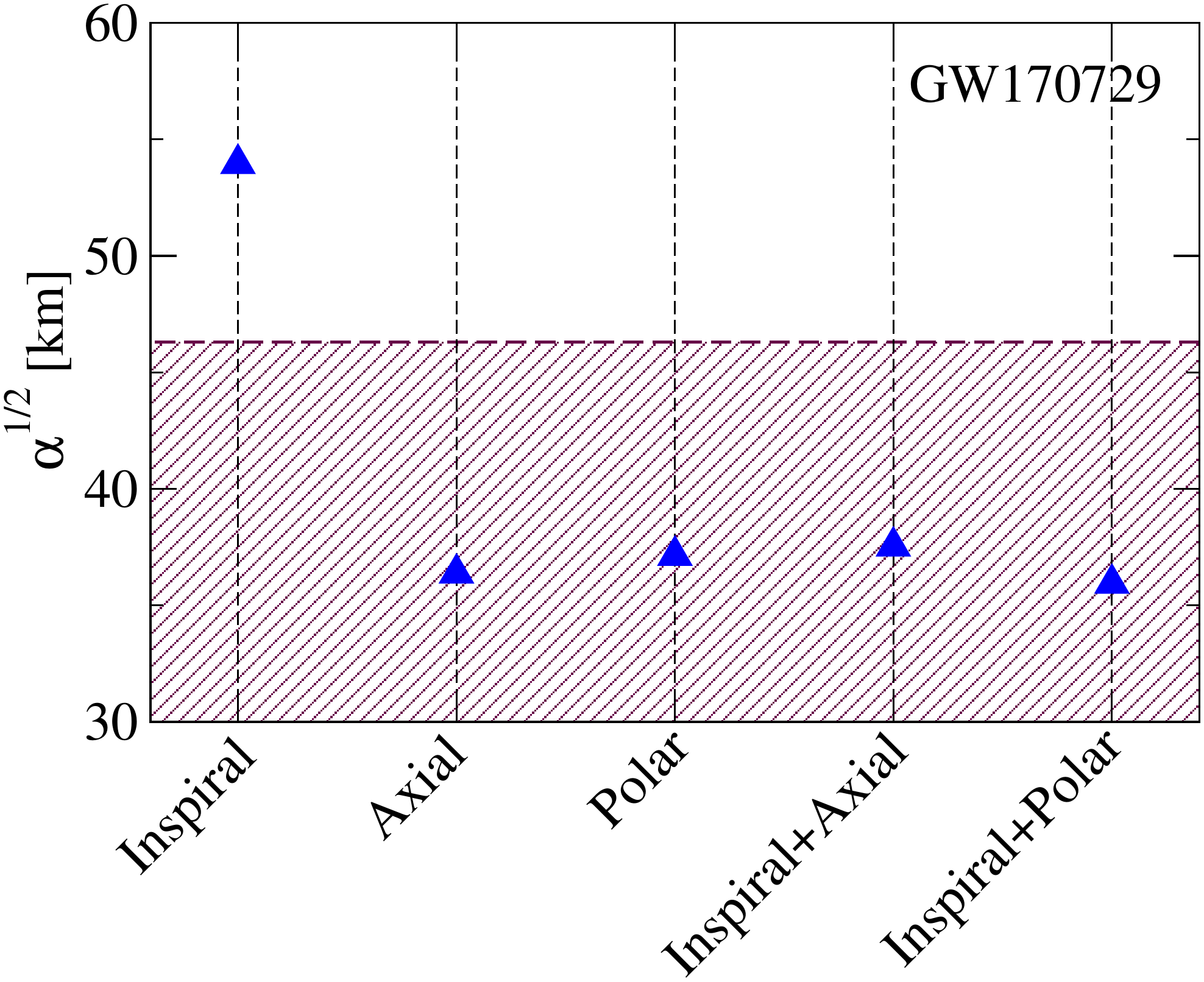}
\caption{
90\%-credible upper bounds on the EdGB parameter $\sqrt{\alpha}$ for the most massive binary BH event detected to date: GW170729.
Such bounds are organized into six categories (represented by the columns in each panel) of EdGB corrections introduced to the GR waveform as discussed in Sec.~\ref{sec:testsOfGR}: inspiral, axial QNMs, polar QNMs, inspiral+axial QNMs, and inspiral+polar QNMs.
Observe the importance of including non-GR effects in the merger-ringdown waveform for massive events, as the small-coupling approximation (valid only in the shaded region) becomes invalid otherwise.
}\label{fig:Param_alpha2}
\end{center}
\end{figure}

\begin{figure}[htb]
\begin{center}
\includegraphics[width=.9\linewidth]{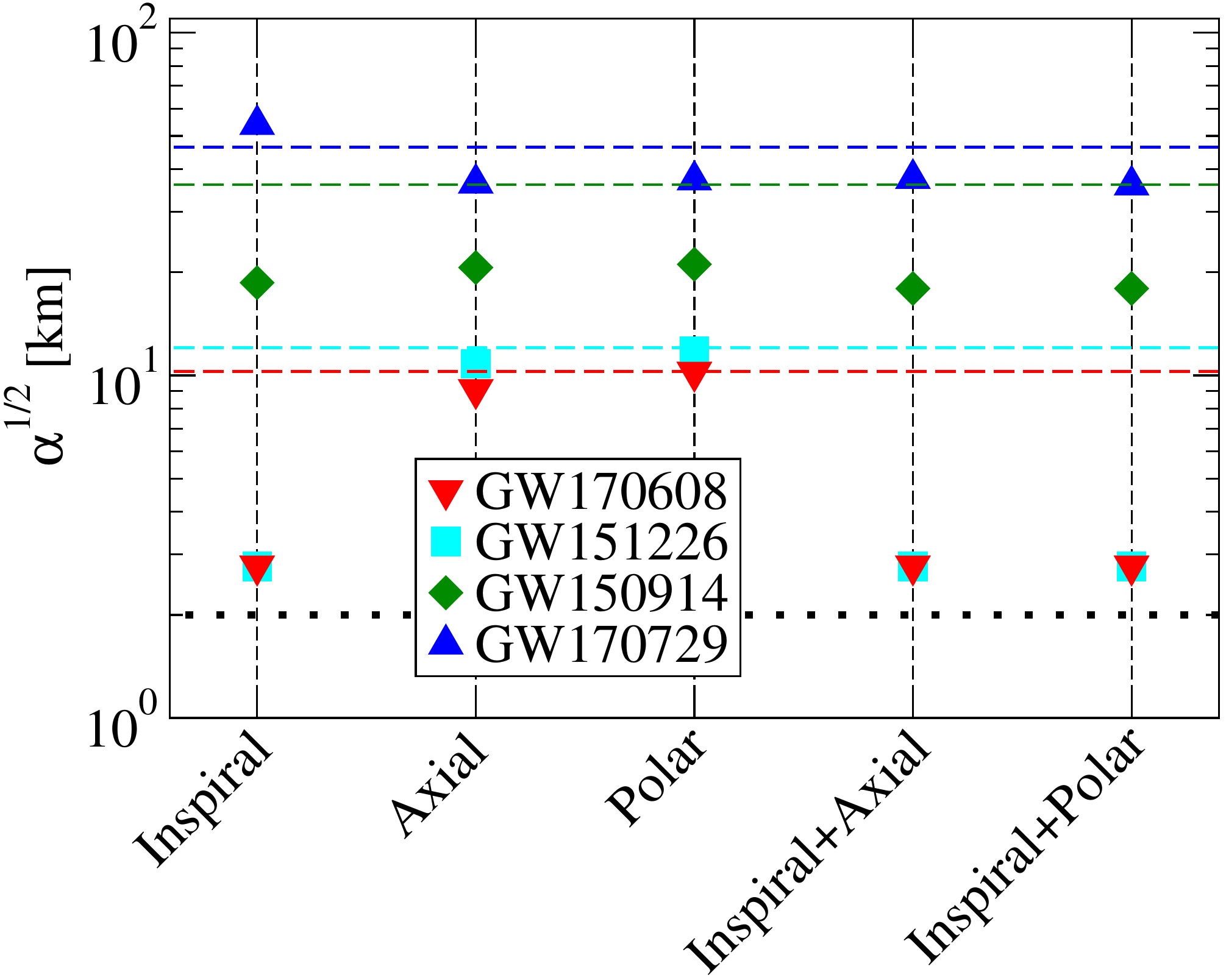}
\caption{
Same as Fig.~\ref{fig:Param_alpha2} but for various GW events GW170608, GW151226, GW150914, and GW170729 (in order of increasing mass).
The dashed horizontal lines represent the small coupling approximation $\zeta \ll 1$ for events of the same color, representing invalid constraints when placed above the corresponding line.
The dotted black horizontal line corresponds to the current constraint of $\sqrt{\alpha}\leq 2$~km.
}\label{fig:Param_alpha1}
\end{center}
\end{figure}

\begin{figure}[htb]
\begin{center}
\includegraphics[width=.9\linewidth]{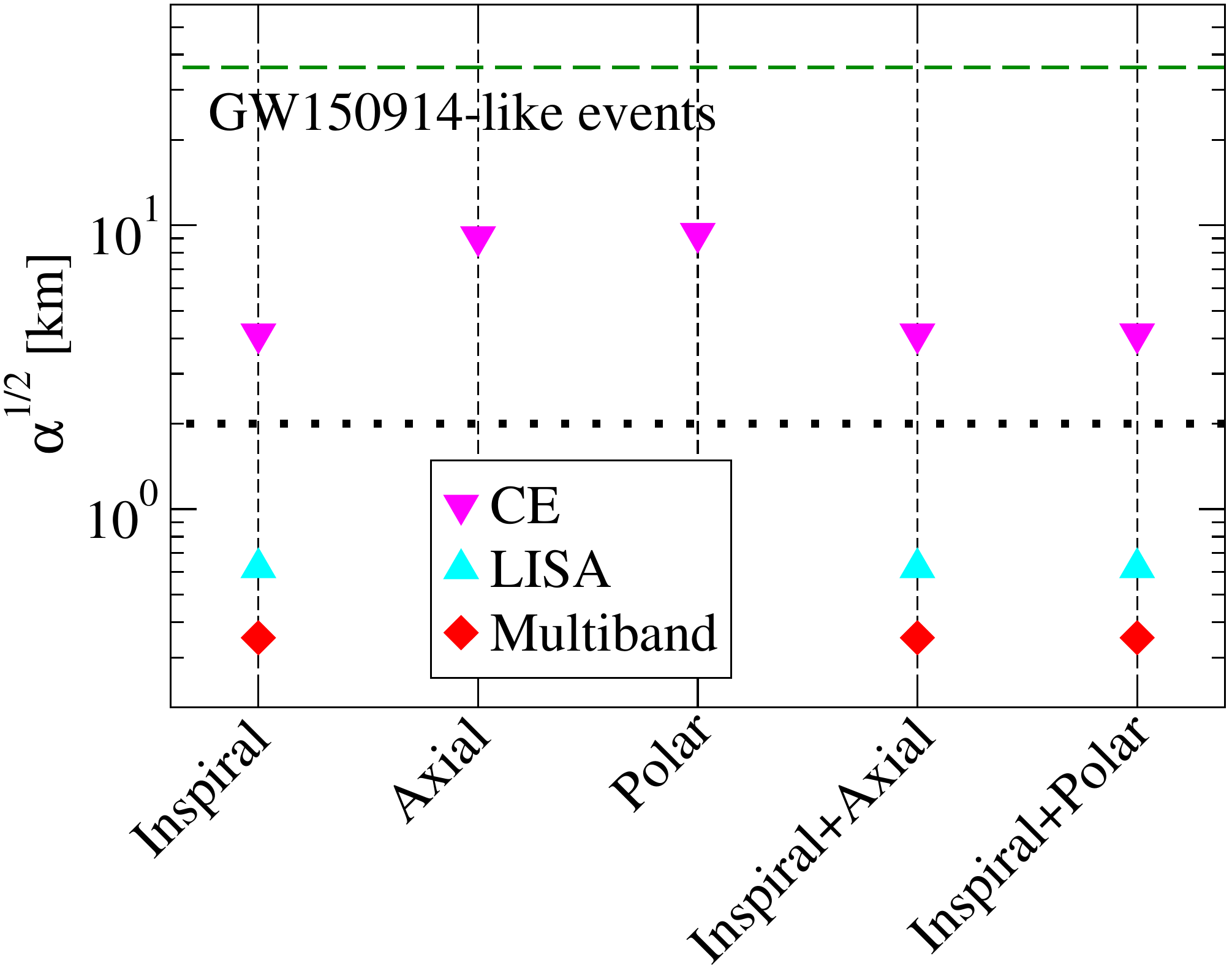}
\caption{
Same as Fig.~\ref{fig:Param_alpha1} but for future GW150914-like events detected by the ground-based detector CE, space-based detector LISA, and the multi-band observation between the two.
We note that no space-based or multiband bounds appear in the axial/polar QNMs columns, because space-based detectors such as LISA can not observe the merger-ringdown effects for GW150914-like events occurring at high frequencies.
}\label{fig:Param_alpha3}
\end{center}
\end{figure}


\section{Results}\label{sec:results}

Now let us discuss the fundamental results obtained in this investigation.
In particular, we present our results  from the GW tests of GR discussed in Sec.~\ref{sec:testsOfGR}, commenting on the estimated constraints given on the EdGB coupling parameter $\alpha$ observed by various GW events, GW detectors, and types of EdGB corrections introduced to the template waveform.
We first perform our main analysis with EdGB corrections to the waveform to $\mathcal{O}(\chi^2)$ in BH spin, followed up by a discussion and demonstration of corrections to $\mathcal{O}(\chi^4)$.

\subsection{$\mathcal{O}(\chi^2)$ corrections to BH spin}\label{sec:2ndOrdSpin}
We start by performing the analysis with EdGB corrections to $\mathcal{O}(\chi^2)$ in BH spin.
Figure~\ref{fig:Param_alpha2} presents the upper bounds on $\sqrt{\alpha}$ from the most massive binary BH event detected to date, GW170729, with various EdGB corrections considered. Observe first that when only the inspiral correction is considered, the bound is beyond the validity of the small coupling approximation, while those become valid once we consider corrections to QNMs. This shows the importance of the latter when constraining EdGB gravity with large mass binaries for which the contribution of the ringdown is larger.

Figure ~\ref{fig:Param_alpha1} displays the upper bounds on $\sqrt{\alpha}$ observed on O1/O2 runs for each GW event considered in this analysis.
We observe several points in regards to this.
Firstly, the smaller total mass events correspond to stronger constraints, as was noted in similar analyses~\cite{Carson_multiBandPRD,Carson_multiBandPRL}.
This is because the expressions in Eqs.~\eqref{eq:alpha_ppE}--\eqref{eq:beta_ppE} minimize $\sqrt{\alpha}$ for both minimal mass ratio \emph{and}, more notably, the individual mass.
Second, we observe that the type of EdGB corrections to the waveform does not strongly affect the two more massive events (GW150914, and GW170729), while the two lighter events (GW170608 and GW151226) observe a deterioration in constraining EdGB gravity when only including the axial/polar QNMs. This is because the fraction of the ringdown portion in the observed waveform becomes larger for larger-mass binaries, and hence QNM corrections become more important for these binaries.
Third, we observe that for more massive events such as GW170729, the inclusion of only inspiral EdGB effects results in an invalid constraint due to violation of the small coupling approximation, as already shown in Fig.~\ref{fig:Param_alpha2}. 
Similar conclusions were made in Ref.~\cite{Carson_QNM_PRL} for massive events, in which the merger-ringdown portion of the gravitational waveform began to make significant contributions to the IMR consistency test, compared to the inspiral portion.

Next we consider the future detectability of EdGB effects in the waveform.
Figure~\ref{fig:Param_alpha3} displays the possible upper bound on $\sqrt{\alpha}$ observed by CE, LISA, and the multiband observation between the two for GW150914-like events, which indeed lie in the multiband detectability region displayed in Ref.~\cite{Carson_multiBandPRD}.
We note that EdGB effects with only axial/polar QNM corrections can not be probed by LISA (thus multiband observations give the same result as CE detections alone) due to its cutoff frequency of $1$ Hz.
We observe that LISA observations alone can improve the ability to probe EdGB gravity by roughly one order of magnitude from CE observations alone, with little difference made by the addition of axial/polar QNM corrections. 
Multiband observations further improve the bound by about a factor of two. Notice also that the LISA and multiband bounds are stronger than current bounds~\cite{Kanti_EdGB,Pani_EdGB,Yagi_EdGB,Witek:2018dmd,Nair_dCSMap,Yamada:2019zrb,Tahura:2019dgr}.
See also Table~\ref{tab:alphas} for a comparison between the $\sqrt{\alpha}$ constraints found in this paper, and the one in Ref.~\cite{Carson_QNM_PRL} through the IMR consistency tests.
In particular, the latter analysis utilized the same EdGB corrections to the gravitational waveform used here, and then tested the consistency between the inspiral and merger-ringdown signals for varying values of $\alpha$.
We find that the bounds from the two analyses are comparable to each other.

\renewcommand{\arraystretch}{1.2}
\begin{table}
\centering
\addvbuffer[12pt 8pt]{\begin{tabular}{c c c}
\multirow{2}{*}{Detector} & $\sqrt{\alpha}$ [km] & $\sqrt{\alpha}$ [km] \\
 & (this paper) &  (IMR consist.~\cite{Carson_QNM_PRL})\\
\hline
aLIGO & $17$  & $15$  \\
CE & $5$  & $8$ \\
LISA & $0.6$  & -- \\
Multiband & $0.3$  & $0.2$  \\
\end{tabular}}
\caption{Comparison between the current and future upper bound on $\sqrt{\alpha}$ obtained in this paper and the IMR consistency tests of GR~\cite{Carson_QNM_PRL}.
Such constraints were formed from GW150914-like events, with the both inspiral and axial QNM EdGB effects included in the waveform template.
Observe how constraints obtained from both tests produce comparable results on the detectability of EdGB effects in the GW signal.
}\label{tab:alphas}
\end{table}

\subsection{$\mathcal{O}(\chi^4)$ corrections to BH spin}\label{sec:4thOrdSpin}
In this section we compute EdGB corrections to the gravitational waveform up to quartic order in BH spin, to check the validity of the slow-rotation approximation to quadratic order in spin used in the previous subsection.
We begin by expanding the expressions for the inspiral dipole radiation and QNM corrections already computed in Sec.~\ref{sec:corrections} to quartic order in BH spin.
Next we compute corrections to $r_\ISCO$,  $E_\text{orb}$ and $L_\text{orb}$ via the EdGB spacetime metric $g_{\alpha\beta}^\EdGB$ found in Ref.~\cite{Maselli:2015tta}, where they computed each element up to 5th order in BH spin $\chi$.
The orbital energy and angular momentum can be obtained from $g_{\alpha\beta}^\EdGB$ by simultaneously solving the equations $V_\text{eff}(r)=0$ and $\frac{d}{dr}V_\text{eff}(r)=0$ for $E_\text{orb}$ and $L_\text{orb}$ with effective potential given by
\begin{equation}
V_\text{eff}(r)=\frac{E_\text{orb}^2g_{\phi\phi}^\EdGB+2E_\text{orb}L_\text{orb}g_{t\phi}^\EdGB+L_\text{orb}^2g_{tt}^\EdGB}{(g_{t\phi}^\EdGB)^2-g_{tt}^\EdGB g_{\phi\phi}^\EdGB}-1.
\end{equation}
Finally, the location of the ISCO is given by  $\frac{d}{dr}E_\text{orb}(r_\ISCO)=0$.

\renewcommand{\arraystretch}{1.2}
\begin{table}
\centering
\addvbuffer[12pt 8pt]{\begin{tabular}{c c c c}
\multirow{2}{*}{GW Event} & $\sqrt{\alpha}$ $(\chi^2)$ & $\sqrt{\alpha}$ $(\chi^4)$ &frac. diff.  \\
  & [km] & [km] & [$\%$]\\
\hline
GW170608~\cite{GW170608} &2.29 & 2.28 & 0.4 \\
GW151226~\cite{GW151226} & 2.76 & 2.75 & 1.1 \\
GW150914~\cite{GW150914} & 17.16 & 17.15 & 0.1 \\
GW170729~\cite{GW170729} & 28.71 & 28.29 & 1.5 \\
\end{tabular}}
\caption{Constraints on $\sqrt{\alpha}$ obtained with EdGB corrections to the waveform up to  quadratic order in BH spin (2nd column), and quartic order in BH spin (3rd column). The last column shows the fractional difference between the two.
We observe that such results agree to within $1.5\%$ in all cases, with the largest difference appearing for the most massive event GW170729.
}\label{tab:4thOrderSpin}
\end{table}

With the above corrections to the entire gravitational waveform to quadratic order in spin, we estimate constraints on EdGB parameter $\sqrt{\alpha}$.
In particular, we compute constraints on $\sqrt{\alpha}$ for each GW event considered in this analysis: GW150914, GW151226, GW170608, and GW170729 as detected on the O2 detector, with non-zero fiducial BH spins.
We compare these results with those of the main analysis, with corrections to only quadratic order in spin.
Table~\ref{tab:4thOrderSpin} presents a comparison between constraints on $\sqrt{\alpha}$ obtained from (i) waveforms with corrections to quadratic order in BH spin, and (ii) to quartic order in BH spin.
We find that such results agree with each other to between $0.1\%$ and $1.5\%$, with the latter resulting from the massive BHs in GW170729, in which spin effects become more important as it seems to have the largest final spin out of the 4 GW events considered here.
Therefore we conclude that the effect of higher-order spin corrections to the gravitational waveform has up to a $\sim1.5\%$ effect on our predictions, which validates our order-of-magnitude estimation presented in this paper including up to quadratic order.

Finally, we consider the effect of including spin effects into the remnant BH QNMs.
For example, in dynamical Chern-Simons (dCS)  gravity, all of the ingredients required to correct the full waveform considered here are available, with the exception of the QNM spin corrections.
Here, we remove all EdGB spin effects to the remnant QNM corrections and compute constraints on $\sqrt{\alpha}$.
We find the constraint to be $27.58\text{ km}$ for GW170729 with which the contribution of the ringdown is most significant out of the 4 GW events considered. Such a constraint agrees very well with those tabulated in Table~\ref{tab:4thOrderSpin} for spin corrections to both $\mathcal{O}(\chi^2)$ and $\mathcal{O}(\chi^4)$.
Therefore, we conclude that spin effects in the remnant BH QNMs make only a negligible impact on constraints on $\sqrt{\alpha}$ in EdGB gravity.


\section{Conclusion and Discussion}\label{sec:conclusion}
EdGB gravity is a proposed scalar-tensor theory of gravity with curvature coupling to the dilaton scalar field.
This string-inspired theory predicts the scalarization of BHs~\cite{Campbell:1991kz,Yunes:2011we,Takahiro,Sotiriou:2014pfa}, calling forth ``fifth force" interactions between orbiting BHs in a binary system and giving rise to scalar dipole radiation that predicts an increased rate of inspiral between them.
In this analysis, we have modeled the resulting EdGB effects throughout various parts of the gravitational waveform, including the inspiral, the characteristic ringdown QNMs, and finally, to the final mass and spin properties of the remnant BH.
With these new tools in hand, we offer predictions on the future detectability of such EdGB effects present in the gravitational waveform. 

We studied the detectability of EdGB effects in an observed GW signal by introducing various combinations of EdGB modifications to the inspiral and merger-ringdown portions of the waveform.
In particular, we discovered that for more massive events such as GW170729, the EdGB merger-ringdown contributions begin to hold high significance.
When only the inspiral corrections to the waveform (as is typically considered) were applied, the small-coupling approximation $\zeta \ll 1$ failed to be upheld. 
Only upon the inclusion of the merger-ringdown corrections does this quantity become satisfied, allowing for valid constraints on $\sqrt{\alpha}$.
We found that future space-based and multiband observations can place bounds that are stronger than current bounds on EdGB gravity.
We also found that the constraints on $\sqrt{\alpha}$ agree well with the recent similar analysis in Ref.~\cite{Carson_QNM_PRL} by the same authors.
In this analysis, the same EdGB corrections to the waveform were made and the inspiral and merger-ringdown signals were tested for consistency with increasing values of $\alpha$ injected into the signal, until they failed to remain consistent with each other.
We improved upon Ref.~\cite{Carson_QNM_PRL} by considering effects at higher order in spin to justify the use of slow-rotation approximation.
We found that such higher-order corrections only change our results up to a maximum of $1.5\%$.
We additionally investigate the effect of spin corrections to the remnant BH QNMs, finding that their inclusion has a negligible impact on parameter estimation.

As noted by Ref.~\cite{Carson_QNM_PRL}, we take note here that the preceding analysis falls short by several assumptions made throughout, however we present it as a first step towards going beyond including the leading correction to the inspiral only or considering QNM corrections only under the BH spectroscopy.
In particular, we offer several caveats to the EdGB waveform corrections displayed in this paper, to be considered as future improvements on similar analyses.
In this investigation, we have made use of the IMRPhenomD GR waveform, which makes several assumptions that hold strong and true in GR.
However, in alternative theories of gravity (like EdGB gravity), such assumptions may fail. 
Firstly, polar and axial QNMs are found to be exactly identical in GR, while those in EdGB gravity are different in general.
Secondly, we would additionally need to consider corrections to the intermediate merger signal, rather than just the inspiral and post-merger ringdown signals.
Finally, in our analysis we consider only leading order PN corrections to the inspiral waveform using the ppE formalism, while higher orders could yet be taken into account.
Having said this, the bounds presented in this paper should serve as valid order-of-magnitude estimates. This is because the correction to the waveform are linearly proportional to $\zeta \propto (\sqrt{\alpha})^4$, which means systematic errors due to mismodeling the waveform is suppressed by a fourth-root power.

Faults such as the ones listed above can be remedied by the full construction of an EdGB (or any non-GR theory) waveform.
Work in this direction is already in progress such as Ref.~\cite{Witek:2018dmd}, where the scalar field dynamics during binary BH mergers have been expressed in EdGB gravity.
Very recently, the EdGB correction to the merger-ringdown waveform from a binary black hole has been computed~\cite{Okounkova:2020rqw}\footnote{See also Refs.~\cite{Okounkova:2017yby,Okounkova:2019dfo,Okounkova:2019zjf} for similar works in dCS gravity.}. 
We plan to compare such numerical-relativity waveforms with the simple analytic model presented here to quantify the validity of the latter.

In the preceding investigation, we considered inspiral-ringdown waveform modifications from the EdGB theory of gravity as one given example. 
Future analyses could, given all the necessary ingredients described above, repeat the entire investigation using any given modified theory of gravity.
By simply knowing the leading PN corrections to the inspiral portion (known for most modified theories of gravity~\cite{Tahura_GdotMap}), corrections to the specific orbital energy $E_\text{orb}$ and angular momentum $L_\text{orb}$, (known for theories such as dCS gravity~\cite{Yagi:2012ya}), and corrections to the QNMs (for dCS gravity, these are only known for non-spinning BHs~\cite{Yunes:2007ss,Cardoso:2009pk,Molina:2010fb}), the simple ``patchwork" analysis presented here could be revisited, without the need for a full non-GR waveform.
Other future avenues include repeating the calculations using a Bayesian analyses.


\acknowledgments
Z.C. and K.Y. acknowledge support from NSF Award PHY-1806776 and the Ed Owens Fund. K.Y. would like to also acknowledge support by a Sloan Foundation Research Fellowship, the COST Action GWverse CA16104 and JSPS KAKENHI Grants No. JP17H06358.


\bibliography{Zack}
\end{document}